# 3D B-mode ultrasound speckle reduction using deep learning for 3D registration applications

Hongliang Li[1], Tal Mezheritsky[1], Liset Vazquez Romaguera[1] and Samuel Kadoury[1,2]
[1] MedICAL laboratory, Polytechnique Montreal, Montreal, Canada
[2] CHUM Research Center, Montreal, Canada

**Abstract.** Ultrasound (US) speckles are granular patterns which can impede image post-processing tasks, such as image segmentation and registration. Conventional filtering approaches are commonly used to remove US speckles, while their main drawback is long run-time in a 3D scenario. Although a few studies were conducted to remove 2D US speckles using deep learning, to our knowledge, there is no study to perform speckle reduction of 3D B-mode US using deep learning. In this study, we propose a 3D dense U-Net model to process 3D US B-mode data from a clinical US system. The model's results were applied to 3D registration. We show that our deep learning framework can obtain similar suppression and mean preservation index (1.066) on speckle reduction when compared to conventional filtering approaches (0.978), while reducing the runtime by two orders of magnitude. Moreover, it is found that the speckle reduction using our deep learning model contributes to improving the 3D registration performance. The mean square error of 3D registration on 3D data using 3D U-Net speckle reduction is reduced by half compared to that with speckles.

**Keywords:** Speckle reduction, Motion prediction, Deep learning, 3D registration, Ultrasound, U-Net

## 1 Introduction

Ultrasound (US) speckles are granular patterns that arise from interference of scattered echoes. US speckles have been exploited in speckle tracking imaging to assess myocardial functions [1], [2], breast cancer [3], [4], [5], [6] and mechanical properties of vascular walls [7], [8], [9], [10], [11]. However, speckles indeed reduce US image contrast, which would undermine other image post-processing tasks, such as image segmentation and registration. Thus, speckles are expected to be removed to improve image contrast.

Speckle reduction can be achieved during US image reconstruction from radio-frequency (RF) channel signals via a compounding strategy. One is able to perform frequency compounding from varying transmission center frequencies, which enables speckle reduction in the frequency domain [12]. Spatial compounding is also used to reduce speckles, in which sub-images at different beam angles are beamformed and compounded to lower speckles [13]. However, speckle reduction using either frequency or spatial compounding will increase system complexity and reduce temporal resolution [14].

Speckle reduction has also been studied in the context of post-processing filtering. Spatial-domain-filtering-based approaches were commonly used to remove US speckles, whose objective is to smooth homogeneous areas while preserving sharp edges. Among them, local filters were first proposed [15], [16], with which the intensity of a pixel in an image with speckle reduction is correlated with its neighboring pixels. Later on, non-local filters were proposed to determine speckle-free pixels in a non-local manner [17], [18], [19]. The filter weights depend not only on spatial distances, but also on pattern similarities. The non-local approaches are considered as the state-of-the-art strategy for ultrasound speckle reduction [20]. While the main drawback is long runtime, which prevents their real-time applications.

Recently, a few studies were conducted to remove US speckles using deep learning. Convolutional neural networks (CNN) were first used to reconstruct speckle-free US images from RF signals [21], [22]. However, RF signals are not accessible for commercial US systems, which impedes its clinical use. Principal component analysis networks (PCANet) [23] was used to extract intrinsic features as inputs to a non-local filter to produce speckle-free US images [20]. Other works investigated whether GANs could perform speckle reduction, in which a generative network was trained to create speckle-free images and a discriminative network differentiate filtered images and outputs from the generative network [24], [25]. Nevertheless, the aforementioned studies processed 2D US datasets which are not demanding computationally.

To our knowledge, there is no study to perform speckle reduction of 3D B-mode US using deep learning models. In this study, we propose a 3D U-Net model to process 3D US B-mode data from a clinical US system. The results without specklar artefacts achieve accurate approximation of speckle-free results from a non-local filter in a real-time manner. We also demonstrated the output of our proposed method improved the registration performance of a fast 3D registration algorithm. The fast and accurate implementation of 3D US speckle reduction allows the possibility of simplifying US real-time applications.

## 2 Methods

### 2.1 Dataset

The dataset consists of liver scans from 24 healthy volunteers who signed written consent. 4D US volume sequences were acquired by a Philips EPIQ7 US system with an X6-1 3D probe under free-breathing. The temporal resolution was 6 Hz. The imaging resolution in axial, lateral and elevational directions were 0.51 mm, 0.81 mm and 1.04 mm, respectively. The total number of volumes was 1286.

Before the training, validation and testing, volumes were preprocessed by mean centering, standard deviation normalization and cropping with a size of 128×128×32 in axial, lateral and elevational views, respectively.

**2.2  Optimized Bayesian non-local means filter**

To obtain ground truth volumes without speckles, each volume was processed by a non-local filter, optimized Bayesian non-local means (OBNLM) filter. Here, we describe briefly the OBNLM filter. More details can be referred to [17]. Specifically, a pixel with speckle reduction, $NL(u)(B_{ik})$, in an image block, $B_{ik}$, is recovered by computing a weighted spatial average of adjacent image blocks, $B_j$, of the current pixel:

$$NL(u)(B_{ik}) = \sum_{B_j} w(B_{ik}, B_j) \mathbf{u}(B_j), \tag{1}$$

where $\mathbf{u}(\cdot)$ is a vector gathering the intensity values of a block, $w(B_{ik}, B_j)$ is a weight for recovering $\mathbf{u}(B_{ik})$ based on the similarity of blocks $B_{ik}$ and $B_j$. In [17], a statistical distance, Pearson distance, $d_P$, was proposed to determine the weight. The Pearson distance assumes the speckle model conforms to following model:

$$u(x) = v(x) + v^{\gamma}(x)\eta(x), \tag{2}$$

where $v(x)$ is the original image, $u(x)$ is the observed image with speckles and $\eta(x) \sim \mathcal{N}(0, \sigma^2)$ is zero-mean Gaussian noise. Unlike other additive Gaussian noise models, this model is more reliable to extract statistics from images as the factor $\gamma$ makes the speckle noise not only depends on US hardware, but also on signal processing during image reconstruction. Considering this speckle model, the Pearson distance is defined by:

$$d_p\left(\mathbf{u}(B_i), \mathbf{u}(B_j)\right) = \sum_{p-1}^{P} \frac{(u^p(B_i) - u^p(B_j))^2}{(u^p)^{2\gamma}(B_j)}, \tag{3}$$

where $P$ is the pixel number in an image block.

In the study, we used an implementation of the OBNLM filter which is publicly available to process original US dataset. The block size is $3 \times 3$ and the number of adjacent image blocks is 3.

**2.3  Model architecture**

The model architecture is described in Fig. 1. The U-Net architecture [26] is the backbone, which was extended into 3D U-Net to allow for 3D input-output. Specifically, the network consisted of an encoder with 4 layers, following a decoder with 4 layers. Each layer includes 3D convolutional operations and ReLU activations. The encoder and decoder are connected by a 3D convolutional layer (the bottom in Fig. 1). The first two dimensions of a volume are halved after passing each encoder layer by a maxpooling operation, while the third dimension is kept unchanged. Likewise, only the first two dimensions are doubled by a transposed convolution operation at the beginning of each decoder layer. Other than skip connections, dense connections were also introduced, which allows the current layer to utilize previous information. This dense U-Net architecture has shown a faster learning on 3D image segmentation [27]. The input is a raw volume with natural speckle. The output is supposed to be the learned speckle-free volume. The speckle reduced volume using the OBNLM filter is used as the target volume for training and evaluation.

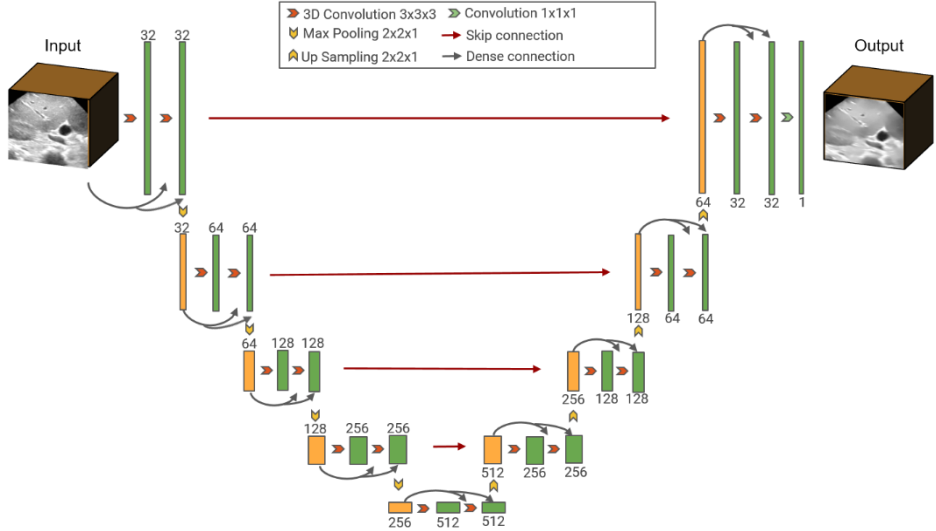

**Fig. 1.** 3D U-Net architecture for speckle reduction. The network is a 4-layer encoder-decoder architecture. Each layer includes 3D convolutional operations and ReLU activations. Skip connections and dense connections are introduced to utilize previous information. The number of features is indicated in the figure.

### 2.4 Training protocol

Volume sequences from 24 volunteers were randomly divided into training, validation and test subsets with a split of 70%, 20% and 10%, respectively. Namely, 842 volumes from 16 volunteers for training, 280 volumes from 5 volunteers for validation and 164 volumes from 3 volunteers are for testing.

The model was deployed with the Keras framework and training was performed on an NVIDIA Quadro P2200 GPU with 16 GB of RAM. The mean square error (MSE), basing on image intensities, between the learned volume without speckles and the ground truth is used as a loss function. Network parameters were optimized using the Adam optimizer [28] with an initial learning rate of $10^{-5}$ and a decay factor of $1.99 \times 10^{-7}$. Due to a limitation of GPU memory, each training batch includes one 3D volume.

### 2.5 Evaluation metrics

**Assessment on speckle reduction:** Speckle suppression and mean preservation index (SMPI) is used to assess the quality of images with speckle reduction [29]. The index, described as follows, avoids overestimations of mean values due to smooth filtering:

$$\text{SMPI} = Q \times \frac{\sqrt{\sigma_r}}{\sqrt{\sigma_o}} \tag{4}$$

where $Q = 1 + |\mu_r - \mu_o|$, $\mu_o$ and $\mu_r$ represent the mean of volume values with and without speckle reduction, respectively. $\sigma_o$ and $\sigma_r$ are the variance of volume values with and without speckle reduction, respectively. A lower SMPI indicates better performance for speckle reduction.

To evaluate computational efficiency, we perform speckle reduction on the test dataset using the trained 3D U-Net on the same GPU and an Intel Xeon E-2244G CPU were recorded. Since the GPU implementation of the OBNLM filter is not available, we only test its runtime on the same CPU. The mean runtime of speckle reduction on per volume is recorded for comparison.

### 2.6   3D registration evaluation

To evaluate the effect of speckle reduction on registration, we train an unsupervised neural network, VoxelMorph, proposed by [30] to perform 3D registration. The registration process minimizes a loss function $\mathcal{L}$:

$$\mathcal{L}(f, m, \phi) = MSE(f, \phi \circ m), \tag{5}$$

where $\phi$ is the registration field and $f$ and $m$ denote the fixed and moving volumes, respectively. The network takes as input the fixed and moving volumes and outputs a dense displacement field which is used to warp the moving volume. The similarity between the fixed and warped volumes is measured with MSE.

Specifically, in a 4D sequence for each volunteer, the volume corresponding to the exhale respiratory phase was visually identified and used as the moving volume. The exhale phase was taken as reference as it is easily reproducible. Then the reference volume was registered to all the other volumes pertaining to the same 4D sequence. The network is trained on three datasets, no speckle reduction, speckle reduction using OBNML and speckle reduction using our 3D U-Net. Each dataset was split into training, validation and testing in the same manner as for the 3D U-Net. The lost in (5) was used to evaluate registration performance.

## 3   Results

**Quality assessment of speckle reduction:** Fig. 2 shows two examples of 3D liver volumes with speckle reduction from two volunteers. One should note that all the results are from the test dataset which is unseen during the training phase. For each example, three slices in coronal, sagittal and transverse views are shown, respectively. As we can see, both images using the OBNML filter and 3D U-Net reduce speckles largely compared with original images. The results using the OBNML and 3D U-Net are similar qualitatively.

Table 1 presents the mean and standard deviation of SMPIs for the test dataset using the OBNLM filter and 3D U-Net. Although the SMPI using the OBNLM filter is smaller than that using the 3D U-Net (0.978 vs 1.066), the small gap between them suggests the 3D U-Net is able to obtain approximated results of speckle reduction comparing to using the OBNLM filter.

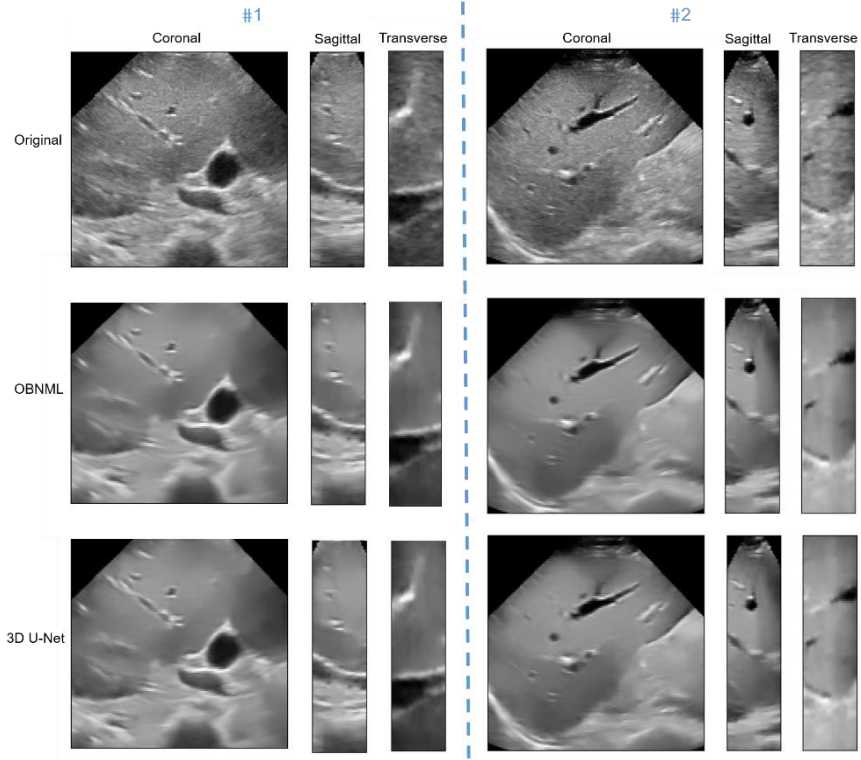

**Fig. 2.** Two slice examples of liver volumes from two volunteers, dividing a dash line. Coronal, sagittal and transverse views are shown, respectively. The first row are original B-mode images. The second row are B-mode images with speckle reduction using the OBNML filter. The third row are B-mode images with speckle reduction using the proposed 3D U-Net.

**Table 1.** Mean and standard deviation of SMPIs of the test dataset

|  | OBNLM | 3D U-Net |
|---|---|---|
| SMPI (mean $\pm$ std) | $0.975 \pm 0.006$ | $1.092 \pm 0.011$ |

**Runtimes:** Table 2 shows the mean runtimes of speckle reduction for each volume of the test dataset using the OBNML filter and 3D U-Net. The runtime using 3D U-Net is reduced by two orders of magnitude with respect to the OBNLM filter runtime.

**Table 2.** Runtime of speckle reduction per volume (seconds)

|     | OBNLM | 3D U-Net |
| --- | --- | --- |
| CPU | 81.1 | 4.0 |
| GPU | - | 0.5 |

**3D registration application:** Table 3 presents the MSE results of 3D registration using VoxelMorph regarding datasets without speckle reduction, with speckle reduction using the OBNLM filter and speckle reduction using 3D U-Net, respectively. The mean MSE of 3D data without speckle using 3D U-Net is reduced by half compared to the data with speckle, which suggests speckle reduction using deep learning can improve registration accuracy.

**Table 3.** Mean and standard of MSE of 3D registration

|     | Original | OBNLM | 3D U-Net |
| --- | --- | --- | --- |
| MSE (mean $\pm$ std) | $0.032 \pm 0.018$ | $0.015 \pm 0.014$ | $0.016 \pm 0.018$ |

Fig. 3 gives two examples of VoxelMorph registration results. The center slice in the coronal view is shown. As seen in the warped results (the third column of each example in Fig. 3), both results using OBNML and 3D U-Net methods reduce registration artifacts as shown in the left part of images.

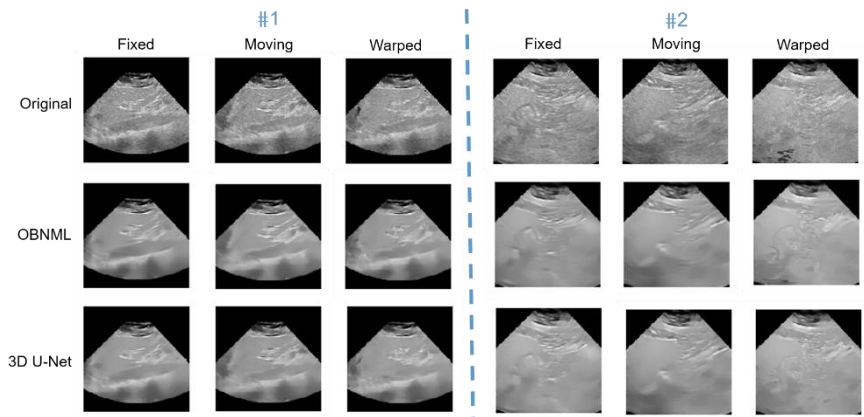

**Fig. 3.** Two slice examples of VoxelMorph registrations, dividing a dash line. The first row shows fixed, moving and warped slices from 3D volumes with speckles. The second row shows fixed, moving and warped slices from 3D volumes without speckles using the OBNML filter. The third row shows fixed, moving and warped slices from 3D volumes without speckle using the proposed 3D U-Net.

## 4 Conclusion

Time-resolved 3D US imaging is emerging. However, with an increasing computation burden, it poses a challenge of reliable 3D post-processing in real-time. In this paper, we attempt to resolve the problem of 3D US speckle reduction and demonstrate its benefits for 3D registration. With a deep learning alternative, we obtained similar performance on speckle reduction compared to a conventional filtering approach, while the computation time was reduced significantly. Applied to a 3D registration problem, it was found that the speckle reduction using our deep learning model contributed to improving registration performance.

**References**

[1] J. D'hooge, E. Konofagou, F. Jamal, A. Heimdal, L. Barrios, B. Bijnens, J. Thoen, F. V. d. Werf, G. Sutherland, and P. Suetens, "Two-dimensional ultrasonic strain rate measurement of the human heart in vivo," *IEEE transactions on ultrasonics, ferroelectrics, and frequency control,* vol. 49, no. 2, pp. 281-286, 2002.

[2] J. Luo, K. Fujikura, S. Homma, and E. E. Konofagou, "Myocardial elastography at both high temporal and spatial resolution for the detection of infarcts," *Ultrasound Med Biol,* vol. 33, no. 8, pp. 1206-1223, Aug, 2007.

[3] E. S. Burnside, T. J. Hall, A. M. Sommer, G. K. Hesley, G. A. Sisney, W. E. Svensson, J. P. Fine, J. Jiang, and N. J. Hangiandreou, "Differentiating benign from malignant solid breast masses with US strain imaging," *Radiology* vol. 245, no. 2, pp. 401-410, 2007.

[4] J. Li, Y. Cui, M. Kadour, and J. A. Noble, "Elasticity reconstruction from displacement and confidence measures of a multi-compressed ultrasound RF sequence," *IEEE Trans Ultrason Ferroelectr Freq Control,* vol. 55, no. 2, pp. 319-326, Feb, 2008.

[5] Y. Zhang, X. Li, S. Li, H. Li, and H. Zheng, "An accurate method of ultrasonic strain estimation under slight tissue compression," in IEEE International Conference on Medical Imaging Physics and Engineering, 2013, pp. 176-181.

[6] Y. Zhang, S. Li, X. Li, H. Li, and H. Zheng, "A method of ultrasonic strain estimation under large tissue compression," in IEEE International Conference on Medical Imaging Physics and Engineering, 2013, pp. 158-161.

[7] H. Li, J. Porée, and G. Cloutier, "A modified affine phase-based estimator for noninvasive vascular ultrasound elastography using coherent plane wave compounding and transverse oscillation imaging," in IEEE International Ultrasonics Symposium (IUS), 2016, pp. 1-4.

[8] H. Li, J. Porée, B. Chayer, M.-H. R. Cardinal, and G. Cloutier, "A global strain estimation algorithm for non-invasive vascular ultrasound elastography," in 2019 IEEE International Ultrasonics Symposium (IUS), 2019, pp. 2210-2213.

[9] H. Li, J. Poree, M. H. Roy Cardinal, and G. Cloutier, "Two-dimensional affine model-based estimators for principal strain vascular ultrasound elastography with compound plane wave and transverse oscillation beamforming," *Ultrasonics,* vol. 91, pp. 77-91, Jan, 2019.

[10] H. Li, B. Chayer, M. H. Roy Cardinal, J. Muijsers, M. van den Hoven, Z. Qin, M. Gesnik, G. Soulez, R. G. P. Lopata, and G. Cloutier, "Investigation of out-of-plane

motion artifacts in 2D noninvasive vascular ultrasound elastography," *Phys Med Biol,* vol. 63, no. 24, pp. 245003, Dec 10, 2018.

[11] H. Li, J. Poree, B. Chayer, M.-H. R. Cardinal, and G. Cloutier, "Parameterized strain estimation for vascular ultrasound elastography with sparse representation," *IEEE Transactions on Medical Imaging*, Jun 25, 2020.

[12] P. A. Magnin, O. T. v. Ramm, and F. L. Thurstone, "Frequency compounding for speckle contrast reduction in phased array images," *Ultrason Imaging,* vol. 4, no. 3, pp. 267-281, 1982.

[13] G. E. Trahey, S. W. Stephen, and O. T. v. Ramm, "Speckle pattern correlation with lateral aperture translation: Experimental results and implications for spatial compounding," *IEEE transactions on ultrasonics, ferroelectrics, and frequency control,* vol. 33, no. 3, pp. 257-264, 1986.

[14] J. Park, J. B. Kang, J. H. Chang, and Y. Yoo, "Speckle reduction techniques in medical ultrasound imaging," *Biomedical Engineering Letters,* vol. 4, no. 1, pp. 32-40, 2014.

[15] Y. Yu, and S. T. Acton, "Speckle reducing anisotropic diffusion," *IEEE Trans Image Process,* vol. 11, no. 11, pp. 1260-1270, 2002.

[16] P. C. Tay, C. D. Garson, S. T. Acton, and J. A. Hossack, "Ultrasound despeckling for contrast enhancement," *IEEE Trans Image Process,* vol. 19, no. 7, pp. 1847-1860, Jul, 2010.

[17] P. Coupe, P. Hellier, C. Kervrann, and C. Barillot, "Nonlocal means-based speckle filtering for ultrasound images," *IEEE Trans Image Process,* vol. 18, no. 10, pp. 2221-2229, Oct, 2009.

[18] J. Yang, J. Fan, D. Ai, X. Wang, Y. Zheng, S. Tang, and Y. Wang, "Local statistics and non-local mean filter for speckle noise reduction in medical ultrasound image," *Neurocomputing,* vol. 195, pp. 88-95, 2016.

[19] T. Wen, J. Gu, L. Li, W. Qin, L. Wang, and Y. Xie, "Nonlocal total-variation-based speckle filtering for ultrasound images," *Ultrason Imaging,* vol. 38, no. 4, pp. 254-275, Jul, 2016.

[20] H. Yu, M. Ding, X. Zhang, and J. Wu, "PCANet based nonlocal means method for speckle noise removal in ultrasound images," *PLoS One,* vol. 13, no. 10, pp. e0205390, 2018.

[21] S. Vedula, O. Senouf, A. M. Bronstein, O. V. Michailovich, and M. Zibulevsky, "Towards CT-quality ultrasound imaging using deep learning," *arXiv preprint arXiv,* vol. 1710.06304, 2017.

[22] D. Hyun, L. L. Brickson, K. T. Looby, and J. J. Dahl, "Beamforming and speckle reduction using neural networks," *IEEE Trans Ultrason Ferroelectr Freq Control,* vol. 66, no. 5, pp. 898-910, May, 2019.

[23] T.-H. Chan, K. Jia, S. Gao, J. Lu, Z. Zeng, and Y. Ma, "PCANet: A simple deep learning baseline for image classification?," *IEEE transactions on image processing,* vol. 24, no. 12, pp. 5017-5032, 2015.

[24] D. Mishra, S. Chaudhury, M. Sarkar, and A. S. Soin, "Ultrasound image enhancement using structure oriented adversarial network," *IEEE Signal Processing Letters,* vol. 25, no. 9, pp. 1349-1353, 2018.

[25] F. Dietrichson, E. Smistad, A. Ostvik, and L. Lovstakken, "Ultrasound speckle reduction using generative adversial networks," in 2018 IEEE International Ultrasonics Symposium (IUS), 2018, pp. 1-4.

[26] O. Ronneberger, P. Fischer, and T. Brox, "U-net: Convolutional networks for biomedical image segmentation." pp. 234-241.

[27] M. Kolařík, R. Burget, V. Uher, K. Říha, and M. Dutta, "Optimized high resolution 3D dense-U-Net network for brain and spine segmentation," *Applied Sciences,* vol. 9, no. 3, 2019.

[28] D. P. Kingma, and J. L. Ba, "Adam: A method for stochastic optimization," *arXiv preprint arXiv,* vol. 1412.6980, 2015.

[29] X. Wang, L. Ge, and L. Xiaojing, "Evaluation of filters for ENVISAT ASAR speckle suppression in pasture area," in ISPAn, 2012, pp. 341-346.

[30] G. Balakrishnan, A. Zhao, M. R. Sabuncu, J. Guttag, and A. V. Dalca, "VoxelMorph: a learning framework for deformable medical image registration," *IEEE Transactions on Medical Imaging,* vol. 38, no. 8, pp. 1788-1800, 2019.